\documentclass[10pt,twocolumn,superscriptaddress,showpacs,pra,aps]{revtex4-1}
\pdfoutput=1

\usepackage{graphicx}   
\usepackage{color}      
\usepackage{units}

\begin{document}

\title{Sisyphus cooling in a continuously loaded trap}

\author{Valentin V. Volchkov} 
\affiliation{5. Physikalisches Institut, Universit\"at Stuttgart, Pfaffenwaldring 57 D-70550 Stuttgart, Germany}
\author{Jahn R\"uhrig} 
\affiliation{5. Physikalisches Institut, Universit\"at Stuttgart, Pfaffenwaldring 57 D-70550 Stuttgart, Germany}
\author{Tilman Pfau}
\affiliation{5. Physikalisches Institut, Universit\"at Stuttgart, Pfaffenwaldring 57 D-70550 Stuttgart, Germany}
\author{Axel Griesmaier}
\affiliation{5. Physikalisches Institut, Universit\"at Stuttgart, Pfaffenwaldring 57 D-70550 Stuttgart, Germany}

\date{\today}

\pacs{31.50.Df, 37.10.De, 37.10.Gh, 37.10.Vz, 37.10.Mn, 67.85.Hj}

\date{\today}

\begin{abstract}

We demonstrate continuous Sisyphus cooling combined with a continuous loading mechanism used to efficiently slow down and accumulate atoms from a guided beam. While the loading itself is based on a single slowing step, applying a radio frequency field forces the atoms to repeat this step many times resulting in a so-called Sisyphus cooling. This extension allows efficient loading and cooling of atoms from a wide range of initial beam conditions.  We study the interplay of the continuous loading and simultaneous Sisyphus cooling in different density regimes. In the case of a low density flux we observe a relative gain in phase-space density of nine orders of magnitude. This makes the presented scheme an ideal tool for reaching collisional densities enabling evaporative cooling - in spite of unfavourable initial conditions.

\end{abstract}

\maketitle

\section{Introduction}

Continuous beams of particles not only constitute the basis of a wide range of fundamental research but also play a major role in high-tech applications, most prominent of them being the traditional caesium beam frequency standard \cite{Ramsey83}. 

The flux defined as the number of particles per second and the phase-space density (PSD) are the most important figures of merit of a particle beam. Since any measurement performed on a beam has a signal-to-noise ratio that directly depends on the flux and the PSD, techniques allowing  to increase the PSD and the number of atoms in the region of interaction are highly desirable and can trigger significant progress. In this spirit, the development of laser cooling and the magneto-optical trap in which atoms from a fast and hot beam are stopped and cooled, increasing the PSD by many orders of magnitude allowed for the implementation of an atomic fountain clock \cite{Kasevich89}. However, this approach can  only be applied to a fraction of possible beams \cite{shuman10,Barry12}. Therefore, alternative approaches like buffer-gas cooling \cite{egorov02}, velocity filtering \cite{Tsuji10} or Stark-deceleration \cite{meerakker05,bethlem99} have been developed to increase the PSD of a beam. In this context, the realization of an atomic diode \cite{price08} that does not rely on laser cooling equally allows stopping and accumulating atoms from a slow, guided beam - eventually leading to Bose-Einstein condensation by subsequent forced evaporative cooling \cite{anoush09,falkenau11}. The dissipation of the directed kinetic energy takes place in form of a single slowing step: the atoms climb up a potential hill, convert most of the kinetic energy into potential energy, which is then removed by changing the internal state of the atoms. Although the idea is general and can be applied to other kinds of beams, the realized experiment strongly relied on a high-flux, slow and transversely cold beam - conditions that are difficult to obtain for an arbitrary beam. Especially, when a single slowing step is not sufficient to dissipate all the kinetic energy, the loading mechanism breaks down. A similar approach demonstrated on a molecular beam \cite{zeppenfeld12} addresses this problem by first trapping the molecules in a confined volume and subsequently repeating the Sisyphus cooling cycles many times. The original idea of Sisyphus cooling involving state dependent potentials and optical pumping was proposed by Pritchard \cite{pritchard83} and has been discussed and implemented in various configurations \cite{Newbury95,Cirac95,Ovchinnikov97,Miller02,Janis05}. 

\begin{figure}
	\centering
	\includegraphics[scale=1]{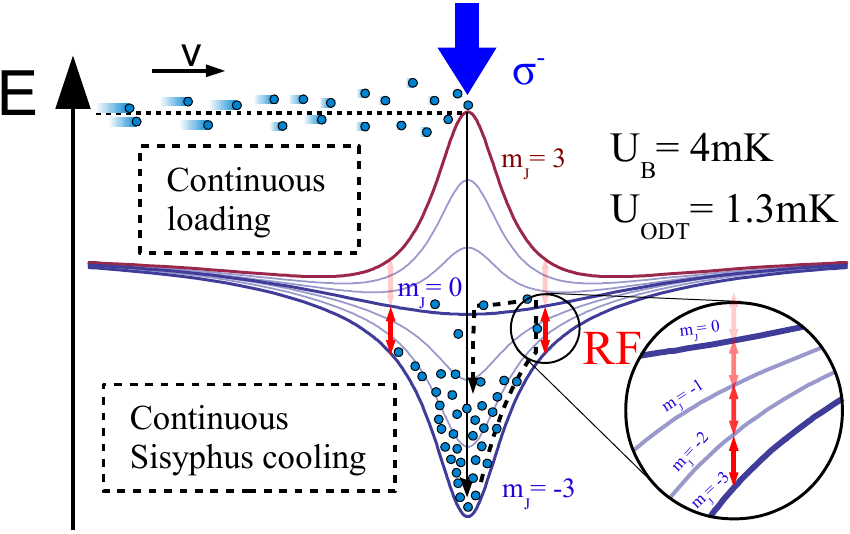}
	\caption{\label{fig:normScheme} Continuous loading and simultaneous cooling mechanism are depicted. The guided atoms in the low-field seeking sub-state $m_J=3$ are slowed down by the repulsive potential, created by a magnetic field barrier in the vicinity of an optical dipole trap. Close to their classical turning point the atoms are pumped into the high-field seeking dark state $m_J=-3$ by optical pumping, denoted by $\sigma^-$. This constitutes the continuous loading mechanism which dissipates a significant part of the longitudinally directed kinetic energy of the atomic beam. The remaining longitudinal energy is further reduced by RF-induced Sisyphus cooling inside the hybrid trap. Each time a trapped atom climbs up the magnetic potential to the regions of magnetic fields where it is resonant to  the radio frequency (enlarged region), its gained potential energy is partly converted into Zeeman energy. The Zeeman energy is eventually dissipated when the atom returns to the centre of the potential where optical pumping light is present.}
\end{figure}

In this paper, we present a combined scheme, in which atoms are continuously loaded from a guided beam and can repeat the Sisyphus cooling cycle to further reduce the kinetic energy as shown in Fig. \ref{fig:normScheme}.  We directly observe the continuous cycling of the Sisyphus cooling and identify different regimes of the combined loading and cooling. As a result, the constraints on the temperature and the flux of the beam necessary to enter collisional regime can be drastically relaxed. We can reduce the flux  and the PSD by several orders of magnitude and still reach favourable conditions for subsequent evaporative cooling. Interestingly, we find that under certain conditions the height of the repulsive potential used to slow the atoms can be reduced as well.

This paper is organized as follows. In Sec. \ref{sec:expsetup} we describe the experimental setup and conditions under which the data was taken. In Sec. \ref{sec:Sisyphus} we show evidence for continuous RF-induced Sisyphus cooling and discuss thee different density regimes, starting from the collisionless and ranging up to density-limited regime. In Sec. \ref{sec:app}, we present the performance of the loading and cooling mechanism under very unfavourable conditions, which constitutes the main result of this paper. An alternative set of experimental parameters for loading and the surprising results are presented in Sec.\ref{sec:noevap}.

\section{Experimental setup} 
\label{sec:expsetup}

The source for our continuous loading experiments is a magnetically guided beam of chromium atoms. It is described and discussed in detail in Refs. \cite{axel09,anoush10,alex07, [{Similar setups with magnetically guided atoms have been also realized by }] Olson:2006.1,*Cren:2002.1}. The relevant properties of the beam are summarized in Table \ref{tab:beam}.
\begin{table}[b]
\begin{ruledtabular}
\centering
\begin{tabular} {lll}
 velocity & 0.4 -- 1 & m/s \\
 transverse temperature & 65 -- 250 & $\mu$K \\
 longitudinal temperature & 100 -- 180 & $\mu$K \\
 flux & $10^5$ -- $10^8$ &Atoms/s \\
\hline
 phase-space density & $10^{-13}$ -- $10^{-8}$ & \\
\end{tabular}
\caption{\label{tab:beam} Range of accessible parameters of the guided atomic beam. Note: the properties shown here are not independent, i.e. reducing the velocity by a factor of two also reduces the flux by two orders of magnitude.}
\end{ruledtabular}
\end{table}

The accumulation of atoms takes place in a hybrid potential, the radial confinement of which is given by an optical dipole trap (ODT), aligned along the guided beam, whereas the axial confinement is dominated by a magnetic trapping potential. Technical details concerning the operation of the atomic diode and the dynamics of the loading process are given in Refs. \cite{anoush09, falkenau11,falkenau12}. 
The oscillation frequencies of the trapped $m_J=-3$ atoms are $\omega_{Rad}=2\pi\times\unit[5.5]{kHz}$ in radial direction and $\omega_{Ax}=2\pi\times\unit[293]{Hz}$ in the axial direction. Atoms in the $m_J=-3$ do not couple to the optical pumping light and experience the strongest axial confinement. One ingredient for the Sisyphus cooling is the fact that transitions to higher lying magnetic sub-states can be driven position-dependently and therefore energy-selectively by a radio frequency (RF) field. We applied a RF field in the range \unit[1--40] {MHz} via a single winding square coil, approximately \unit[2]{cm} away from the atoms. The coil with the dimension of \unit[4]{cm} was driven by a $\mathrm{P_{max}=\unit[250]{mW}}$ frequency generator without any impedance matching. For a sufficiently low RF-power high energy atoms end up in states with $1>m_J>-3$ that are still trapped but are less confined in axial direction. Therefore their potential energy is converted into Zeeman energy.  In order to close the Sisyphus cooling cycle, the atoms scatter optical pumping light near the bottom of the trap and return into the $m_J=-3$ dark-state, as shown in Fig. \ref{fig:normScheme}.

\section{Evidence for Sisyphus cooling}
\label{sec:Sisyphus}

In this section, we present experimental data that confirm the multiple iterations of the Sisyphus cycles as the major reason for the observed cooling effect and rule out RF-induced evaporation in axial direction. For this purpose, we study the cooling dynamics at low densities at timescales shorter than collision times. In this collisionless regime, the cooling effect is a single-atom process and can be  simulated by numerical methods. Increased amount of scattered light from the centre of the continuously loaded trap is observed on fluorescence images and constitutes another signature for the simultaneous Sisyphus cooling. We study the performance of the cooling depending on the applied frequency and power. We relate the results to the fluorescence images. In steady state, the thermalization plays an important role, combined with RF-assisted Sisyphus cooling, the temperature is significantly reduced - far below the collisionless regime. Finally, we discuss how entering the density-limited regime reduces the efficiency of the cooling mechanism.

\subsection{Collisionless regime}
\label{sec:collless}

\begin{figure*}
	\centering
	\includegraphics[scale=1]{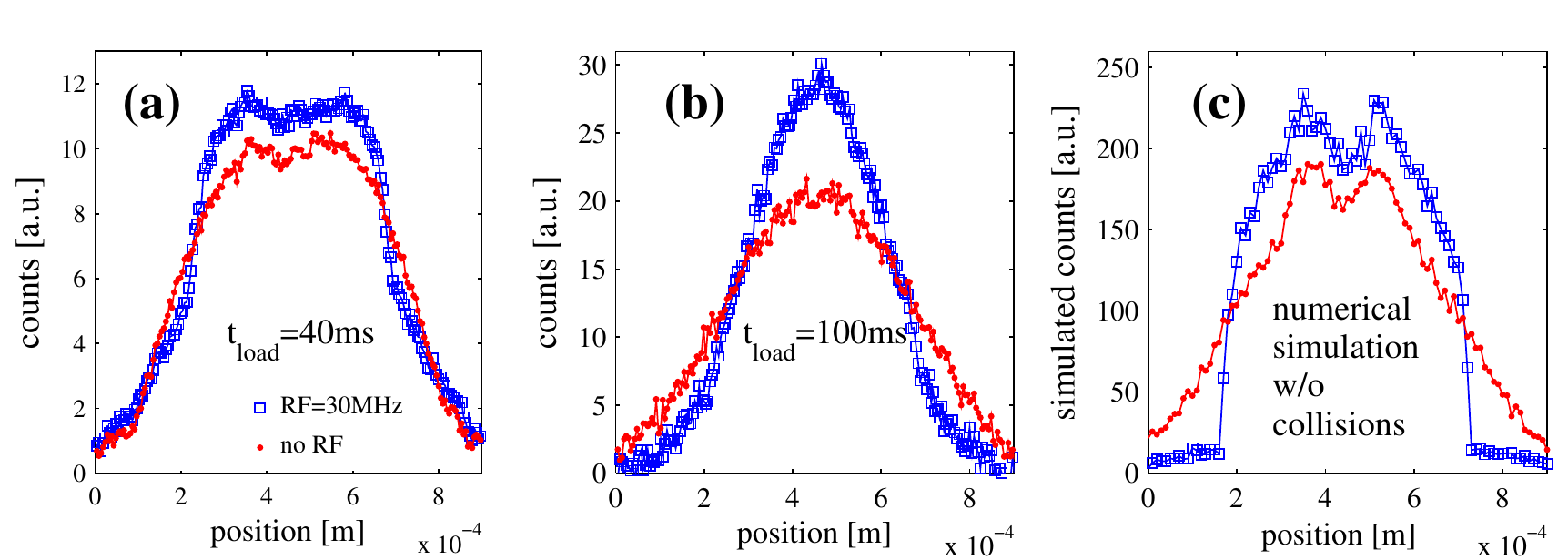}
\caption{\label{fig:fluo}   Axial profiles of TOF images of atoms. They represent the momentum distribution convoluted with the in-trap position distribution. Data represented by squares corresponds to optimized RF-cooling, whereas the dots represent usual loading without RF. The data in (a) demonstrate the effect of RF in the collisionless regime, that is, after \unit[40]{ms} of loading into the trap.  Both profiles correspond to out-of-equilibrium distributions, however one can clearly see that applying RF transfers atoms from the high energy wings into the central region. This effect can be reproduced by numerical simulation of a large number of non-interaction atoms and is shown in (c). Eventually, after a loading time of \unit[100]{ms} the trapped atoms begin to thermalize and the profiles in (b) take the shape of a Gaussian.}
\end{figure*}

The RF-assisted Sisyphus cooling as shown in Fig. \ref{fig:normScheme} is a single particle effect. As long as trapped atoms do not collide, each atom loaded into the trap performs his own cooling cycles. The axial kinetic energy is reduced in each Sisyphus cycle until the RF-region can no longer be reached, this leads to a cut-off in the momentum distribution. For the same reason we also expect a truncated position distribution. Profiles (a)-(c) in Fig. \ref{fig:fluo} show the distributions of atoms after a time-of-flight (TOF) of \unit[300]{$\mu$s}, integrated along the two radial directions. Since the TOF does not exceed the oscillation period, what we see is the convolution of the position distribution with the momentum distribution. Graph (a) on Fig. \ref{fig:fluo} shows profiles after a loading time of \unit[40]{ms} with and without RF. We can clearly see that high energy atoms from the wings of the distributions are transferred to the centre when RF is applied. Profiles in graph (c) originate from a Monte-Carlo simulation of the loading and cooling process and show the same signature. In order to estimate the timescales involved in the loading and cooling process, we varied the loading time between \unit[10]{ms} and \unit[100]{ms}. At short loading time ($\mathrm{t_{load}}=$\unit[10]{ms}) the profiles are identical, we therefore infer that the timescale for cooling lies on the order of \unit[20-30]{ms}. At $\mathrm{t_{load}}=$\unit[100]{ms} the profiles shown in (b) assume the shape of a Gaussian which suggests collisional thermalization of the sample. In the presence of RF the temperature is significantly lower than without RF. The the total number of atoms during the collisionless time is not reduced by RF  which implies the absence of axial evaporation. Moreover, numerical simulations at very high RF powers show that the loading mechanism breaks down because atoms are transferred adiabatically from $m_J=3$ to $m_J=-3$ prior to being optically pumped and trapped. 

\subsection{Steady state properties}
\label{sec:steadystate}

In order to describe the steady state properties of the continuously loaded trap, it is convenient to relate the thermal energy $k_BT$ of the sample to the depth of the optical dipole trap $U_{ODT}=k_B\times\unit[1.3]{mK} $ by defining $\eta_{ODT}=U_{ODT}/k_BT$, where $k_B$ is the Boltzmann constant. $\eta_{ODT}$ determines the speed and the efficiency of the evaporation during the loading \cite{falkenau12}. Similarly, we introduce $\eta_{RF}=U_{RF}/k_BT$ as the ratio of the thermal energy and the RF threshold energy $U_{RF}$ for the stretched dark state $m_J=-3$. At the bottom of the hybrid trap the magnetic field has its maximum in axial direction, where it amounts to $B_0=$\unit[12]{G}. The energy splitting between the neighbouring magnetic sub-states is given by $\Delta E_{RF}=2\mu_B B$ and the frequency necessary to drive transitions is given by $f_{RF}=\Delta E_{RF}/h$. For $B_0$ we have $f_{RF0}=\unit[33.8]{MHz}$. Hence, the magnetic potential energy of the $m_J=-3$ atoms at the location of the RF can be written as $U_{RF}=3h\times(f_{RF0}-f_{RF})=3h\times\Delta f_{RF}$. In units of temperature  the frequency difference $\Delta f_{RF}=$\unit[1]{MHz} corresponds to $U_{RF}=k_B\times\unit[144]{\mu K} $. 

The steady state is reached when the number of atoms loaded into the trap is saturated and the radial and axial motional degrees of freedom are in equilibrium via collisional thermalization. The loading time necessary to reach steady state varies as a function of flux between a few hundreds of milliseconds and twenty seconds.  As opposed to collisionless regime, RF-induced Sisyphus cooling takes place despite $\eta_{RF}>1$, as the high energy tail of the energy distribution is constantly refilled by collisions. This situation is comparable to (plain) evaporation, however, in our case, atoms promoted to a state above the RF-threshold are not lost from the system.

\begin{figure}
	\centering
		\includegraphics[scale=1]{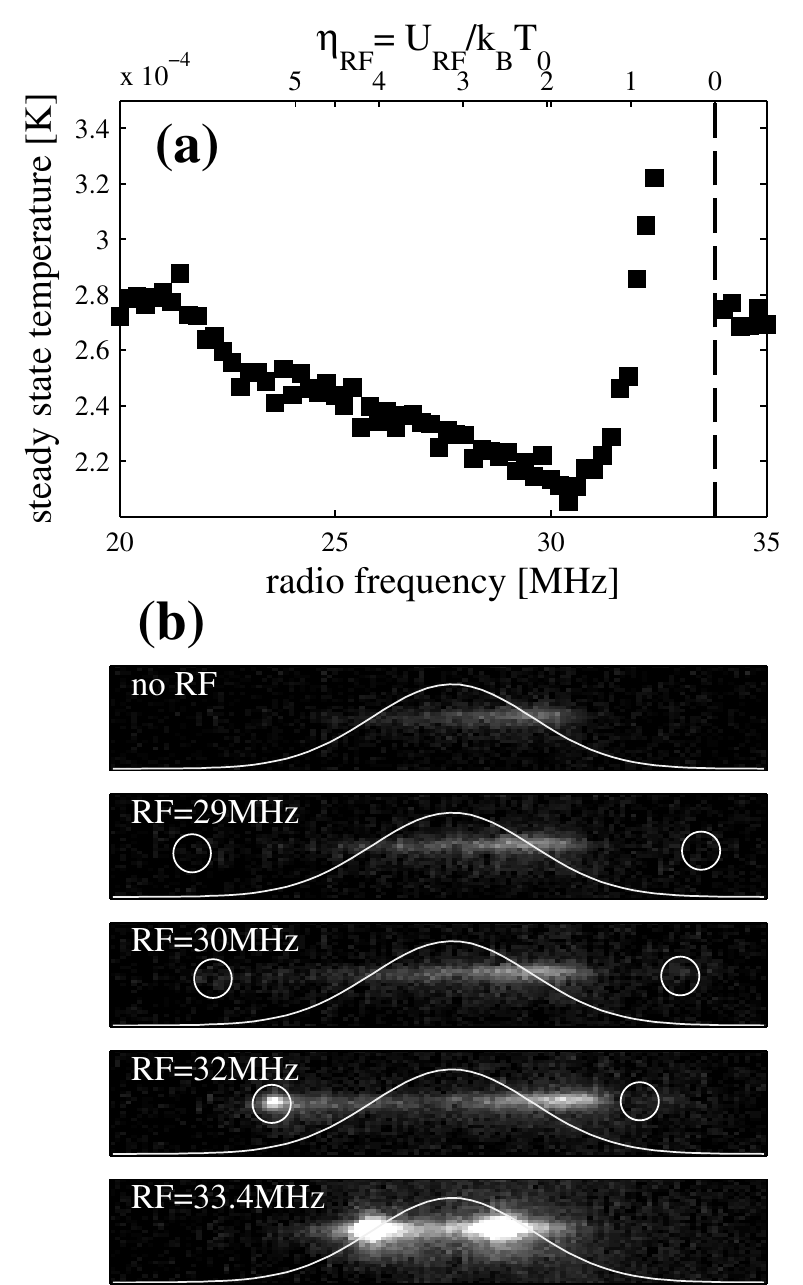}
		\caption{\label{fig:TvsRF}  The temperature of the atomic sample in steady state is plotted versus the applied radio frequency in (a). The RF threshold $\eta_{RF}$ axis (top) is calculated with respect to initial temperature $T_0=$\unit[280]{$\mu$K}.  The vertical dashed line indicates the position of the trap bottom at the frequency of \unit[33.8]{MHz}, above which the radio frequency field shows no effect. Pictures in (b) show the fluorescence of the atoms being loaded into and cooled in the hybrid trap in steady state. The optical pumping beam enters the depicted region from top and crosses the atomic trap in the centre. The position and the shape of the optical pumping beam are indicated by the Gaussian profile with a $1/e^2$ waist of $\unit[140]{\mu m}$ (white curve). When no RF is applied (top picture), the fluorescence corresponds to light scattered during the loading process only. The applied radio frequency manifests itself in increased fluorescence in the centre of the trap (see Fig. \ref{fig:RFamplitude}), as well as at the position where RF transitions are driven, highlighted by white circles. }
\end{figure}

We characterize the RF-induced cooling by performing loading experiments at different frequencies and powers. Figure \ref{fig:TvsRF} (a) shows the steady state temperature of the loaded trap as a function of the applied radio frequency. Although one would expect that for frequencies below \unit[25]{MHz} ($\eta_{RF}>\eta_{ODT}$) evaporation in radial direction should strongly suppress Sisyphus cooling by removing hot atoms, one has to keep in mind that new atoms are being continuously added to the trap. As it is pointed out in Ref. \cite{falkenau11} and  as we have seen in the collisionless case above, those new atoms have much larger kinetic energy in axial  than in radial direction before they thermalize and redistribute that energy in the cloud. Therefore the cooling effect already sets in for frequencies above \unit[22]{MHz}.  The lowest temperature is found around \unit[30]{MHz}. We observe a strong increase of temperature when the radio frequency approaches the trap centre. We can explain this behaviour by taking the finite size of the optical pumping beam into account. Figure \ref{fig:TvsRF} (b) shows a series of fluorescence pictures taken for increasing frequencies. The extent of the optical pumping beam is illustrated (to scale) by the white Gaussian profile. The onset of the heating coincides with the frequency where the location of the RF-resonance (white circles) enters the spacial extent of the pumping light beam.

\begin{figure}
	\centering
		\includegraphics[scale=1]{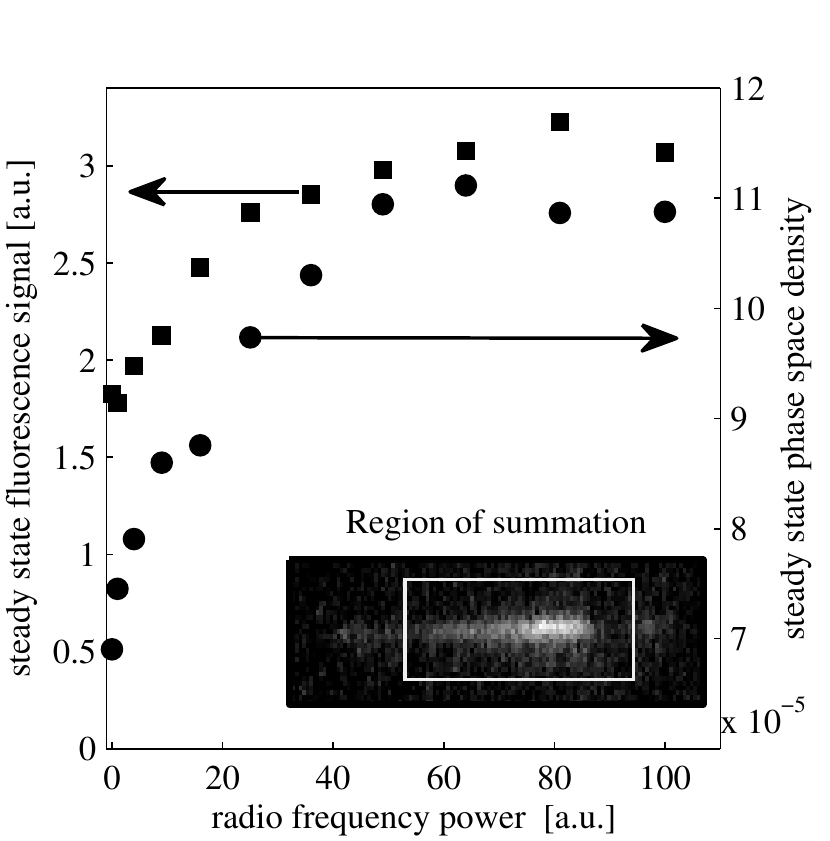}
		\caption{\label{fig:RFamplitude} Saturation of the PSD and fluorescence for increasing RF power. The values of the fluorescence signal (squares, left y-axis) are obtained from a summation of the indicated region (see inset), while the  phase-space density (circles, right y-axis) is extracted from separate absorption pictures of the atomic sample.}
\end{figure}

We study the performance of the Sisyphus cooling in terms of gain in PSD as a function of the RF field power and present the results in Fig. \ref{fig:RFamplitude}. For each power setting we determine the phase-space density from an absorption picture. In addition to that we take a fluorescence picture giving access to the relative number of scattered photons. We consider only the light collected from the centre of the cloud as the fluorescence signal, as indicated by the region of summation in the inset of Fig. \ref{fig:RFamplitude}. The simultaneous increase of phase-space density with the fluorescence signal proves the occurrence of the Sisyphus cooling. We observe saturation of PSD at large RF powers. The dynamics of the cooling take place in the axial direction only, meaning that every time an atom interacts with the RF field it cannot avoid the optical pumping light on its oscillation back. For this reason, the fluorescence signal is directly linked to the probability of an atom to undergo a transition into another Zeeman sub-state. The strongest cooling effect is achieved when atoms are transferred into the $m_J=0$ sub-state (see Fig. \ref{fig:normScheme}.) However, the state in which the atom ends up 
cannot be addressed without also population the neighbouring states.  The saturation of the phase-space density and the fluorescence signal indicate that transitions into the untrapped states begin to play a role and  the loading mechanism itself gets disturbed.

\subsection{Density-limited regime}
\label{sec:hydro}

\begin{figure}
	\centering
		\includegraphics[scale=1]{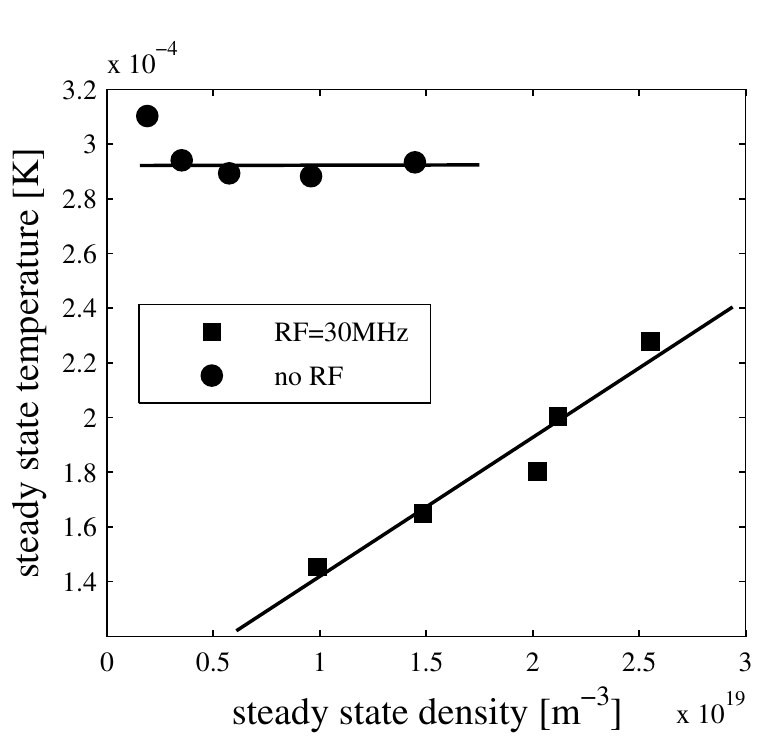}
		\caption{\label{fig:hydro} The flux of the atomic beam is varied, leading to different steady state densities of trapped atoms and corresponding temperatures. In the case of loading without RF(circles), the temperature is approximately constant. In the case of RF-assisted loading (squares) the temperature is significantly lower. However, it increases linearly with density (lines are guides to the eye).}
\end{figure}

In the case of the continuous loading without additional Sisyphus cooling, the limiting factor are elastic two-body collisions and the resulting evaporation of atoms during the loading \cite{falkenau12}.  Ultimately, the same is true for the Sisyphus-cooled loading mechanism, nevertheless, the density can be significantly increased in the presence of RF-induced cooling. To demonstrate and study this effect, we vary the steady state density by reducing the flux of the atomic beam, leading to different steady state numbers of trapped atoms and corresponding temperatures. Figure \ref{fig:hydro} shows that without RF-induced cooling the steady state temperature of the fully loaded trap does not depend on the final density. Given the same atomic beam conditions and with optimized Sisyphus cooling, the temperature shows a linear dependence on the steady state density in the investigated regime. As a consequence, the PSD-gain of the RF-induced Sisyphus cooling is density dependent, which will be discussed in the next section. 

There are at least three possible mechanisms for the density dependent temperature increase. Firstly, the heating through reabsorption: as a matter of fact, the resonant optical density along the short axis of the cloud varies between 10 and 30. As a result, photons scattered in the optical pumping process are reabsorbed multiple times, leading to a heating depending on the density \cite{Lewenstein98}. However, this cannot be the dominant reason, since scattered photons originating from the loading mechanism would also lead to significant heating, which is not observed in the experiment \footnote{One can estimate from Fig. \ref{fig:RFamplitude} that the amount of scattered light from loading and Sisyphus cooling are approximately the same.}. Secondly, one may also expect reduced efficiency of the Sisyphus cooling due to collisional recapture.  As discussed above, for $k_BT<U_{RF}$ the Sisyphus cooling mechanism relies on collisional production of high energy atoms. In the case where the collision rate exceeds the oscillation frequency, high energy atoms likely undergo further collisions during a single oscillation period and may fall below the RF-threshold again. For given experimental conditions, the collision rate rises to $\Gamma_{coll}\geq\unit[1500]{s^{-1}}\approx\omega_{Ax}$, entering the so-called hydrodynamic regime in the axial direction. Finally, excited state collisions have been identified in optical cooling of dense chromium samples to be responsible for strong losses rather than heating. Here again, this loss or heating mechanism is not particular to the Sisyphus-cooling and should be equally observable for the optical pumping in the loading process without additional RF field.

Lastly, due to the fact that the above mentioned effects cannot be studied separately, a thorough investigation of the limiting mechanism goes beyond the scope of this paper.

\section{Applications: achieving collisional regime from hot and dilute atomic beam }
\label{sec:app}

\begin{figure}
	\centering
		\includegraphics[scale=1]{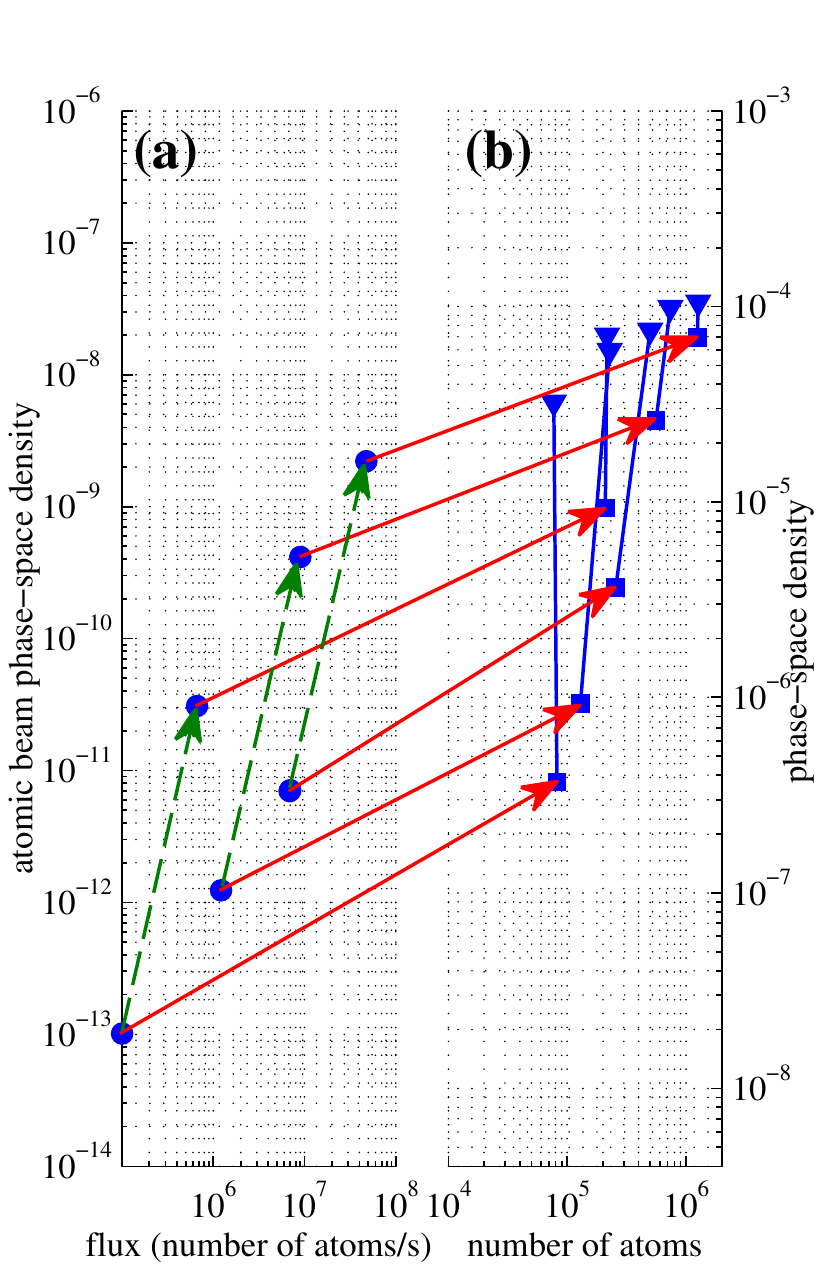}
		\caption{\label{fig:PSDvsN}   Continuous loading and cooling results (b) for different starting conditions (a) are summarized. The atomic beam in (a) is characterized by phase-space density and the flux. The data points denoted by circles correspond to different atomic beams arriving in the loading region. The flux of the atomic beam can be varied, as well as the radial temperature. The dashed arrows therefore indicate the phase-space density increase from continuous radial cooling of the atomic beam. The arrows from (a) to (b) represent the continuous loading process. The squares in (b) correspond to the phase-space density and the number of atoms in the continuously loaded trap in steady state without the RF field, whereas the triangles denote the steady state numbers with simultaneous RF-induced Sisyphus cooling.}
\end{figure}

According to the density-dependent steady state temperature discussed in the previous section, the PSD gain of the Sisyphus-cooling drops for increasing flux and resulting higher steady state density. Therefore, the achievable steady state PSD saturates regardless of the increasing flux in the density-limited regime. The potential of the presented continuous Sisyphus cooling therefore lies in the low-density regime, especially when the phase-space density of the atomic beam is so low that the continuous loading scheme consisting of a single Sisyphus step fails to reach collisional density. 

In order to demonstrate the capacity of the Sisyphus cooling to enable evaporative cooling for unfavorable starting conditions, we performed a series of experiments with decreasing flux and phase-space density of the atomic beam. Hereby, we implement experimental conditions relevant for beams that cannot be cooled radially and suffer from strong reduction of flux when for example velocity filtering techniques are applied \cite{Tsuji10}.

In Fig. \ref{fig:PSDvsN}, the continuous loading is depicted by the solid arrows from (a) to (b), connecting the various atomic beam conditions in (a) to the steady state phase-space density versus the number of atoms in the fully loaded trap (squares) in (b). The gain  given by the (vertical) lines in (b) corresponds to the simultaneous continuous Sisyphus cooling (triangles). The gain in PSD reaches two orders of magnitude when the density limitations of the Sisyphus cooling as discussed above do not play a role. Here we would like to emphasize that even when the radial cooling of the atomic beam is not used \cite{anoush10} and the flux is reduced by almost three orders of magnitude (lowest, left-most data point in Fig. \ref{fig:PSDvsN} (a)) the total overall gain relative to the atomic beam amounts to over nine orders of magnitude. For this case the steady state temperature of the RF-enhanced loading scheme being $T=\unit[150]{\mu K}$ corresponds to $\eta_{RF}\approx 3.6$ and implies a high elastic collision rate, so that only the hottest atoms participate in the RF-induced Sisyphus cooling. We have experimentally checked that the subsequent evaporative cooling of the fully loaded trap is equally  efficient due to a large initial truncation parameter $\eta_{ODT}>9$. The ability to reach this regime from an almost arbitrary beam  \footnote{The only constraint for this scheme is that the longitudinal velocity of the beam has to match the height of the barrier. In Sec. \ref{sec:noevap} this constraint is further relaxed.} constitutes the main result of this paper.

\begin{figure}
	\centering
	\includegraphics[scale=1]{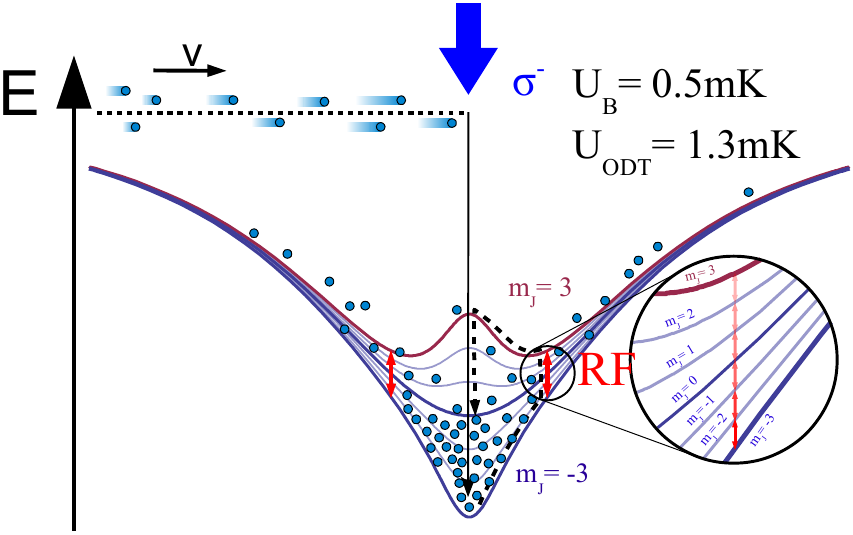}
	\caption{\label{fig:lowBarrier}   Alternative loading scheme for extra slow beams. Compared to scheme  in Fig. \ref{fig:normScheme}, the velocity of the incoming atomic beam is reduced to v=\unit[0.4]{m/s}.  At the same time, the barrier height is on the order of one half of the kinetic energy of the atomic beam. The depth of the ODT is kept constant. As seen in the scheme, the atoms are first accelerated by the ODT before being slowed down by the repulsive barrier. Subsequently, optical pumping dissipates in total an energy corresponding to twice the barrier height. In this alternative scheme, the atom remains trapped radially  \textit{and axially} in the ODT and can repeat the optical pumping process when RF is applied.}
\end{figure}

\section{Evaporationless regime}
\label{sec:noevap}

\begin{figure}
	\centering
	\includegraphics[scale=1]{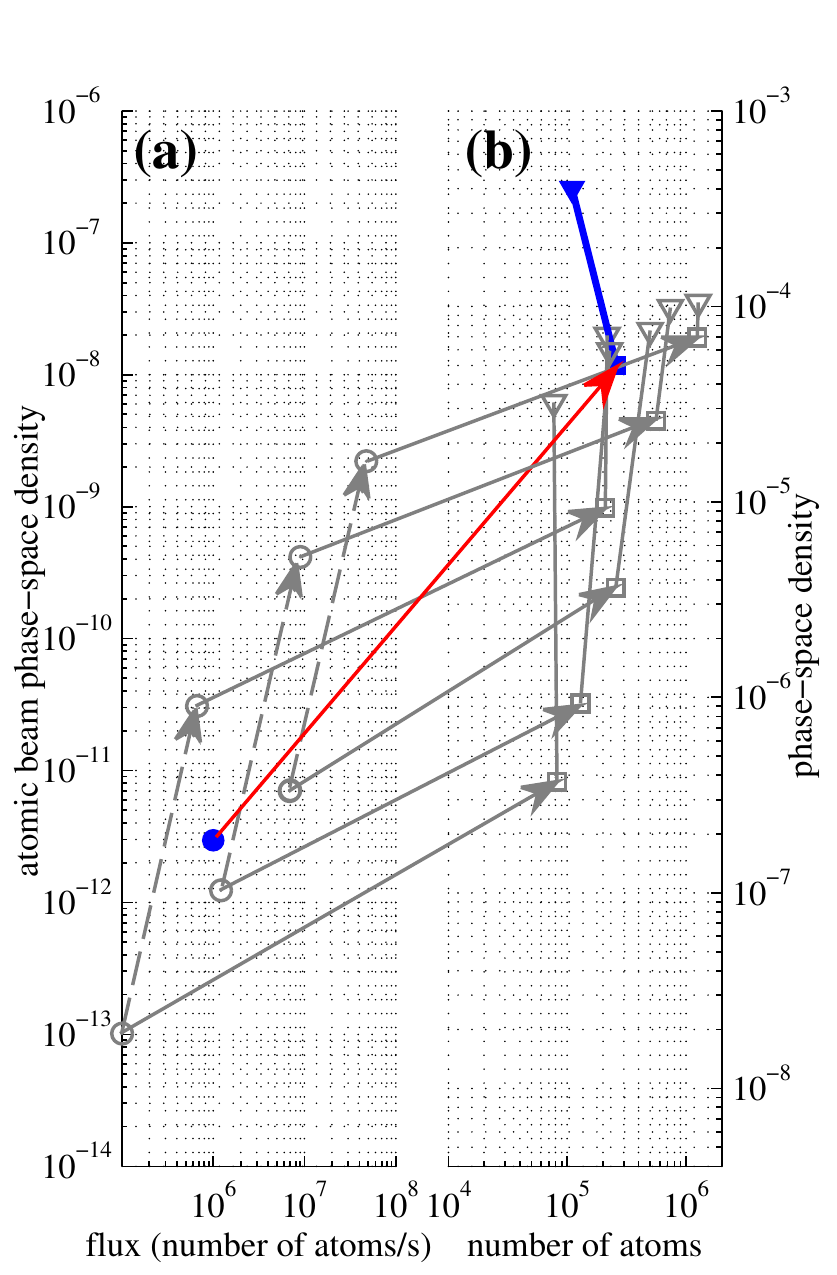}
	\caption{\label{fig:lowBarrierPSD_Plot}  The performance of the alternative loading scheme is shown analogously to Fig. \ref{fig:PSDvsN}. The data from Fig. \ref{fig:PSDvsN} is shown in gray for comparison. The data point with the highest PSD in (b) corresponds to an atomic sample with a steady state temperature of $T=50\mu K$, at which evaporation is strongly suppressed.   }
\end{figure}

In our original proposal \cite{anoush09} and up to now the overall potential is chosen such that the incoming atoms in the low-field seeking $m_J=3$ Zeeman state have their turning points as close to the potential maximum as possible \cite{anoush09,falkenau11}. This condition is vital in the limit when the initial kinetic energy $\mathrm{E_{Kin}}$ is much larger than the depth of the optical dipole trap $\mathrm{U_{ODT}}$, since any excess energy can be redistributed in radial direction, leading to a fast removal (spilling) of atoms from the ODT. In particular, $\mathrm{U_{ODT}}$ needs to be taken into account by adding it to the initial kinetic energy in order to determine the necessary barrier height $\mathrm{U_B=E_{Kin}+U_{ODT}}$. In the opposite limit, when the initial kinetic energy is less than one sixth of $\mathrm{U_{ODT}}$, accumulation can take place by means of evaporation only \cite{Roos:2003.2}.

In this section we explore the intermediate regime when the initial directed kinetic energy is on the order of $\mathrm{U_{ODT}}$. In this case, the purpose of the barrier is no longer to stop the atoms at the top of the potential, but to remove sufficient energy so that the atoms remain trapped in the combined potential for the high-field seeking state, which results in $\mathrm{U_B\geq 1/2E_{Kin}}$. 
Indeed, the significant advantage of the lower potential barrier is that it is technically less demanding to create. We lowered the magnetic field barrier such that the trap bottom is found to be at $f_{RF0}=\unit[3.7]{MHz}$. With RF field at $f_{RF}=\unit[2.5]{MHz}$ we obtain $U_{RF}=\unit[173]{\mu K}$. Figure \ref{fig:lowBarrier} illustrates this alternative scheme. We reduce the velocity to $v=\unit[0.4]{m/s}$ in order to fulfil the conditions mentioned above. As a consequence the flux drops by more than two orders of magnitude and the radial cooling cannot be used \cite{anoush10}. This results in a slow but very dilute and radially hot atomic beam, as shown in Fig. \ref{fig:lowBarrierPSD_Plot} (a). The loading mechanism in Fig. \ref{fig:lowBarrier} still works by dissipating the Zeeman-energy via optical pumping in a Sisyphus step. However, the remaining kinetic energy can go up to $\mathrm{U_{ODT}}$ and has to be either removed by collisional thermalization and evaporation or by repeating the dissipative Sisyphus step when a RF field is applied. In contrast to Fig. \ref{fig:normScheme}, the atoms cannot be transferred into an axially untrapped state. The results of this loading scheme are presented in Fig. \ref{fig:lowBarrierPSD_Plot} (b). Similarly to Fig. \ref{fig:PSDvsN}, the steady state for RF-enhanced loading is denoted by the filled triangle, whereas plain loading without the RF field is denoted by the filled square. The steady state temperatures are $\mathrm{T_{RF}=\unit[50]{\mu K}}$ and $\mathrm{T=\unit[120]{\mu K}}$ respectively. We obtain the highest steady state PSD of $4\cdot 10^{-4}$ with $10^5$ atoms in the trap when RF-induced Sisyphus cooling is present. This PSD is significantly higher than in all previously described experiments in spite of the low initial PSD of the atomic beam!

We also observe that under such loading conditions the number of atoms is reduced by a factor of 2.5 when the RF field is applied. As outlined in \cite{falkenau12}, the steady state number of atoms is governed by the loading rate, the background loss rate and the evaporation rate. Here, the loading rate as well as the background loss rate do not depend on the presence of the RF field. The rate of evaporation depends on the temperature and is strongly suppressed for $\mathrm{T_{RF}=\unit[50]{\mu K}}$, corresponding to  $\eta_{ODT}\geq 26$. From these arguments we must assume that an additional loss process is involved when the continuous loading is accompanied by RF-induced Sisyphus cooling. Since the density does not exceed $\unit[10^{19}]{m^{-3}}$ we do not expect any losses associated with very high densities. In case of multiple optical pumping events experienced by each atom the existence of metastable states may become important. The decay of the used $^7$P$_3$ excited state has a branching ratio of $1:10^3$ to a metastable state, which can be safely neglected in case of a small number of optical pumping processes. In the present loading scheme, however,  the number of scattered photons per atom can become very large when large excess kinetic energy has to be dissipated by numerous Sisyphus steps. In order to investigate this assumption, additional repumping lasers are required. This will be the subject of future work.

\section{Conclusion}
\label{sec:outlook}

We have shown that the implementation of the continuous RF-induced Sisyphus cooling significantly widens the range of initial beam conditions that can be transferred via continuous loading into an optical dipole trap. In particular, atoms loaded with the assistance of Sisyphus cooling are efficiently brought into collisional regime, from which subsequent evaporative cooling can be used. We have demonstrated the generality of this approach by varying the transverse temperature and the flux of the initial beam, covering thereby more than five orders of magnitude in the initial PSD of the beam. We have also shown that with the techniques presented in this paper, the velocity can be reduced at the cost of flux and still a high PSD can be reached in the continuously loaded trap. In conclusion, the application of the Sisyphus cooling in continuous beam experiments facilitates the loading into a conservative trap by relaxing the requirements on the properties of the beam or by reducing the technical constraints. This opens direct access to evaporative cooling for species that cannot be brought to high densities otherwise.

On the one hand, once the trap reaches steady state, evaporative cooling can be used in the next step to increase the phase-space density. On the other hand, RF-induced Sisyphus cooling following the completion of a loading cycle can be viewed as a lossless alternative to evaporative cooling and is the subject of current investigation.

\section*{Acknowledgments}

We acknowledge funding from the Deutsche Forschungsgemeinschaft (DFG) under contract number PF381/14-1. 


\begin{thebibliography}{28}%
\makeatletter
\providecommand \@ifxundefined [1]{%
 \@ifx{#1\undefined}
}%
\providecommand \@ifnum [1]{%
 \ifnum #1\expandafter \@firstoftwo
 \else \expandafter \@secondoftwo
 \fi
}%
\providecommand \@ifx [1]{%
 \ifx #1\expandafter \@firstoftwo
 \else \expandafter \@secondoftwo
 \fi
}%
\providecommand \natexlab [1]{#1}%
\providecommand \enquote  [1]{``#1''}%
\providecommand \bibnamefont  [1]{#1}%
\providecommand \bibfnamefont [1]{#1}%
\providecommand \citenamefont [1]{#1}%
\providecommand \href@noop [0]{\@secondoftwo}%
\providecommand \href [0]{\begingroup \@sanitize@url \@href}%
\providecommand \@href[1]{\@@startlink{#1}\@@href}%
\providecommand \@@href[1]{\endgroup#1\@@endlink}%
\providecommand \@sanitize@url [0]{\catcode `\\12\catcode `\$12\catcode
  `\&12\catcode `\#12\catcode `\^12\catcode `\_12\catcode `\%12\relax}%
\providecommand \@@startlink[1]{}%
\providecommand \@@endlink[0]{}%
\providecommand \url  [0]{\begingroup\@sanitize@url \@url }%
\providecommand \@url [1]{\endgroup\@href {#1}{\urlprefix }}%
\providecommand \urlprefix  [0]{URL }%
\providecommand \Eprint [0]{\href }%
\providecommand \doibase [0]{http://dx.doi.org/}%
\providecommand \selectlanguage [0]{\@gobble}%
\providecommand \bibinfo  [0]{\@secondoftwo}%
\providecommand \bibfield  [0]{\@secondoftwo}%
\providecommand \translation [1]{[#1]}%
\providecommand \BibitemOpen [0]{}%
\providecommand \bibitemStop [0]{}%
\providecommand \bibitemNoStop [0]{.\EOS\space}%
\providecommand \EOS [0]{\spacefactor3000\relax}%
\providecommand \BibitemShut  [1]{\csname bibitem#1\endcsname}%
\let\auto@bib@innerbib\@empty
\bibitem [{\citenamefont {Ramsey}(1983)}]{Ramsey83}%
  \BibitemOpen
  \bibfield  {author} {\bibinfo {author} {\bibfnamefont {N.~F.}\ \bibnamefont
  {Ramsey}},\ }\href {http://tf.nist.gov/general/pdf/1916.pdf} {\bibfield
  {journal} {\bibinfo  {journal} {JOURNAL OF RESEARCH of the National Bureau of
  Standards}\ }\textbf {\bibinfo {volume} {88}},\ \bibinfo {pages} {301}
  (\bibinfo {year} {1983})}\BibitemShut {NoStop}%
\bibitem [{\citenamefont {Kasevich}\ \emph {et~al.}(1989)\citenamefont
  {Kasevich}, \citenamefont {Riis}, \citenamefont {Chu},\ and\ \citenamefont
  {DeVoe}}]{Kasevich89}%
  \BibitemOpen
  \bibfield  {author} {\bibinfo {author} {\bibfnamefont {M.~A.}\ \bibnamefont
  {Kasevich}}, \bibinfo {author} {\bibfnamefont {E.}~\bibnamefont {Riis}},
  \bibinfo {author} {\bibfnamefont {S.}~\bibnamefont {Chu}}, \ and\ \bibinfo
  {author} {\bibfnamefont {R.~G.}\ \bibnamefont {DeVoe}},\ }\href {\doibase
  10.1364/ON.15.12.000031} {\bibfield  {journal} {\bibinfo  {journal} {Optics
  News}\ }\textbf {\bibinfo {volume} {15}},\ \bibinfo {pages} {31} (\bibinfo
  {year} {1989})}\BibitemShut {NoStop}%
\bibitem [{\citenamefont {Shuman}\ \emph {et~al.}(2010)\citenamefont {Shuman},
  \citenamefont {Barry},\ and\ \citenamefont {DeMille}}]{shuman10}%
  \BibitemOpen
  \bibfield  {author} {\bibinfo {author} {\bibfnamefont {E.~S.}\ \bibnamefont
  {Shuman}}, \bibinfo {author} {\bibfnamefont {J.~F.}\ \bibnamefont {Barry}}, \
  and\ \bibinfo {author} {\bibfnamefont {D.}~\bibnamefont {DeMille}},\
  }\href@noop {} {\bibfield  {journal} {\bibinfo  {journal} {Nature (London)}\
  ,\ \bibinfo {pages} {820}} (\bibinfo {year} {2010})}\BibitemShut {NoStop}%
\bibitem [{\citenamefont {Barry}\ \emph {et~al.}(2012)\citenamefont {Barry},
  \citenamefont {Shuman}, \citenamefont {Norrgard},\ and\ \citenamefont
  {DeMille}}]{Barry12}%
  \BibitemOpen
  \bibfield  {author} {\bibinfo {author} {\bibfnamefont {J.~F.}\ \bibnamefont
  {Barry}}, \bibinfo {author} {\bibfnamefont {E.~S.}\ \bibnamefont {Shuman}},
  \bibinfo {author} {\bibfnamefont {E.~B.}\ \bibnamefont {Norrgard}}, \ and\
  \bibinfo {author} {\bibfnamefont {D.}~\bibnamefont {DeMille}},\ }\href
  {\doibase 10.1103/PhysRevLett.108.103002} {\bibfield  {journal} {\bibinfo
  {journal} {Phys. Rev. Lett.}\ }\textbf {\bibinfo {volume} {108}},\ \bibinfo
  {pages} {103002} (\bibinfo {year} {2012})}\BibitemShut {NoStop}%
\bibitem [{\citenamefont {Egorov}\ \emph {et~al.}(2002)\citenamefont {Egorov},
  \citenamefont {Lahaye}, \citenamefont {Sch\"ollkopf}, \citenamefont
  {Friedrich},\ and\ \citenamefont {Doyle}}]{egorov02}%
  \BibitemOpen
  \bibfield  {author} {\bibinfo {author} {\bibfnamefont {D.}~\bibnamefont
  {Egorov}}, \bibinfo {author} {\bibfnamefont {T.}~\bibnamefont {Lahaye}},
  \bibinfo {author} {\bibfnamefont {W.}~\bibnamefont {Sch\"ollkopf}}, \bibinfo
  {author} {\bibfnamefont {B.}~\bibnamefont {Friedrich}}, \ and\ \bibinfo
  {author} {\bibfnamefont {J.~M.}\ \bibnamefont {Doyle}},\ }\href {\doibase
  10.1103/PhysRevA.66.043401} {\bibfield  {journal} {\bibinfo  {journal} {Phys.
  Rev. A}\ }\textbf {\bibinfo {volume} {66}},\ \bibinfo {pages} {043401}
  (\bibinfo {year} {2002})}\BibitemShut {NoStop}%
\bibitem [{\citenamefont {Tsuji}\ \emph {et~al.}(2010)\citenamefont {Tsuji},
  \citenamefont {Sekiguchi}, \citenamefont {Mori}, \citenamefont {Momose},\
  and\ \citenamefont {Kanamori}}]{Tsuji10}%
  \BibitemOpen
  \bibfield  {author} {\bibinfo {author} {\bibfnamefont {H.}~\bibnamefont
  {Tsuji}}, \bibinfo {author} {\bibfnamefont {T.}~\bibnamefont {Sekiguchi}},
  \bibinfo {author} {\bibfnamefont {T.}~\bibnamefont {Mori}}, \bibinfo {author}
  {\bibfnamefont {T.}~\bibnamefont {Momose}}, \ and\ \bibinfo {author}
  {\bibfnamefont {H.}~\bibnamefont {Kanamori}},\ }\href
  {http://stacks.iop.org/0953-4075/43/i=9/a=095202} {\bibfield  {journal}
  {\bibinfo  {journal} {Journal of Physics B: Atomic, Molecular and Optical
  Physics}\ }\textbf {\bibinfo {volume} {43}},\ \bibinfo {pages} {095202}
  (\bibinfo {year} {2010})}\BibitemShut {NoStop}%
\bibitem [{\citenamefont {van~de Meerakker}\ \emph {et~al.}(2005)\citenamefont
  {van~de Meerakker}, \citenamefont {Smeets}, \citenamefont {Vanhaecke},
  \citenamefont {Jongma},\ and\ \citenamefont {Meijer}}]{meerakker05}%
  \BibitemOpen
  \bibfield  {author} {\bibinfo {author} {\bibfnamefont {S.~Y.~T.}\
  \bibnamefont {van~de Meerakker}}, \bibinfo {author} {\bibfnamefont
  {P.~H.~M.}\ \bibnamefont {Smeets}}, \bibinfo {author} {\bibfnamefont
  {N.}~\bibnamefont {Vanhaecke}}, \bibinfo {author} {\bibfnamefont {R.~T.}\
  \bibnamefont {Jongma}}, \ and\ \bibinfo {author} {\bibfnamefont
  {G.}~\bibnamefont {Meijer}},\ }\href {\doibase 10.1103/PhysRevLett.94.023004}
  {\bibfield  {journal} {\bibinfo  {journal} {Phys. Rev. Lett.}\ }\textbf
  {\bibinfo {volume} {94}},\ \bibinfo {pages} {023004} (\bibinfo {year}
  {2005})}\BibitemShut {NoStop}%
\bibitem [{\citenamefont {Bethlem}\ \emph {et~al.}(1999)\citenamefont
  {Bethlem}, \citenamefont {Berden},\ and\ \citenamefont {Meijer}}]{bethlem99}%
  \BibitemOpen
  \bibfield  {author} {\bibinfo {author} {\bibfnamefont {H.~L.}\ \bibnamefont
  {Bethlem}}, \bibinfo {author} {\bibfnamefont {G.}~\bibnamefont {Berden}}, \
  and\ \bibinfo {author} {\bibfnamefont {G.}~\bibnamefont {Meijer}},\ }\href
  {\doibase 10.1103/PhysRevLett.83.1558} {\bibfield  {journal} {\bibinfo
  {journal} {Phys. Rev. Lett.}\ }\textbf {\bibinfo {volume} {83}},\ \bibinfo
  {pages} {1558} (\bibinfo {year} {1999})}\BibitemShut {NoStop}%
\bibitem [{\citenamefont {Price}\ \emph {et~al.}(2008)\citenamefont {Price},
  \citenamefont {Bannerman}, \citenamefont {Viering}, \citenamefont
  {Narevicius},\ and\ \citenamefont {Raizen}}]{price08}%
  \BibitemOpen
  \bibfield  {author} {\bibinfo {author} {\bibfnamefont {G.~N.}\ \bibnamefont
  {Price}}, \bibinfo {author} {\bibfnamefont {S.~T.}\ \bibnamefont
  {Bannerman}}, \bibinfo {author} {\bibfnamefont {K.}~\bibnamefont {Viering}},
  \bibinfo {author} {\bibfnamefont {E.}~\bibnamefont {Narevicius}}, \ and\
  \bibinfo {author} {\bibfnamefont {M.~G.}\ \bibnamefont {Raizen}},\ }\href
  {\doibase 10.1103/PhysRevLett.100.093004} {\bibfield  {journal} {\bibinfo
  {journal} {Phys. Rev. Lett.}\ }\textbf {\bibinfo {volume} {100}},\ \bibinfo
  {pages} {093004} (\bibinfo {year} {2008})}\BibitemShut {NoStop}%
\bibitem [{\citenamefont {Aghajani-Talesh}\ \emph {et~al.}(2009)\citenamefont
  {Aghajani-Talesh}, \citenamefont {Falkenau}, \citenamefont {Griesmaier},\
  and\ \citenamefont {Pfau}}]{anoush09}%
  \BibitemOpen
  \bibfield  {author} {\bibinfo {author} {\bibfnamefont {A.}~\bibnamefont
  {Aghajani-Talesh}}, \bibinfo {author} {\bibfnamefont {M.}~\bibnamefont
  {Falkenau}}, \bibinfo {author} {\bibfnamefont {A.}~\bibnamefont
  {Griesmaier}}, \ and\ \bibinfo {author} {\bibfnamefont {T.}~\bibnamefont
  {Pfau}},\ }\href@noop {} {\bibfield  {journal} {\bibinfo  {journal} {Journal
  of Physics B: Atomic Molecular and Optical Physics}\ }\textbf {\bibinfo
  {volume} {42}},\ \bibinfo {pages} {245302} (\bibinfo {year}
  {2009})}\BibitemShut {NoStop}%
\bibitem [{\citenamefont {Falkenau}\ \emph {et~al.}(2011)\citenamefont
  {Falkenau}, \citenamefont {Volchkov}, \citenamefont {R\"uhrig}, \citenamefont
  {Griesmaier},\ and\ \citenamefont {Pfau}}]{falkenau11}%
  \BibitemOpen
  \bibfield  {author} {\bibinfo {author} {\bibfnamefont {M.}~\bibnamefont
  {Falkenau}}, \bibinfo {author} {\bibfnamefont {V.~V.}\ \bibnamefont
  {Volchkov}}, \bibinfo {author} {\bibfnamefont {J.}~\bibnamefont {R\"uhrig}},
  \bibinfo {author} {\bibfnamefont {A.}~\bibnamefont {Griesmaier}}, \ and\
  \bibinfo {author} {\bibfnamefont {T.}~\bibnamefont {Pfau}},\ }\href {\doibase
  10.1103/PhysRevLett.106.163002} {\bibfield  {journal} {\bibinfo  {journal}
  {Phys. Rev. Lett.}\ }\textbf {\bibinfo {volume} {106}},\ \bibinfo {pages}
  {163002} (\bibinfo {year} {2011})}\BibitemShut {NoStop}%
\bibitem [{\citenamefont {Zeppenfeld}\ \emph {et~al.}(2012)\citenamefont
  {Zeppenfeld}, \citenamefont {Englert}, \citenamefont {Gl\"{o}öckner},
  \citenamefont {Prehn}, \citenamefont {Mielenz}, \citenamefont {Sommer},
  \citenamefont {van Buuren}, \citenamefont {Motsch},\ and\ \citenamefont
  {Rempe}}]{zeppenfeld12}%
  \BibitemOpen
  \bibfield  {author} {\bibinfo {author} {\bibfnamefont {M.}~\bibnamefont
  {Zeppenfeld}}, \bibinfo {author} {\bibfnamefont {B.~G.~U.}\ \bibnamefont
  {Englert}}, \bibinfo {author} {\bibfnamefont {R.}~\bibnamefont
  {Gl\"{o}öckner}}, \bibinfo {author} {\bibfnamefont {A.}~\bibnamefont
  {Prehn}}, \bibinfo {author} {\bibfnamefont {M.}~\bibnamefont {Mielenz}},
  \bibinfo {author} {\bibfnamefont {C.}~\bibnamefont {Sommer}}, \bibinfo
  {author} {\bibfnamefont {L.~D.}\ \bibnamefont {van Buuren}}, \bibinfo
  {author} {\bibfnamefont {M.}~\bibnamefont {Motsch}}, \ and\ \bibinfo {author}
  {\bibfnamefont {G.}~\bibnamefont {Rempe}},\ }\href {\doibase
  10.1038/nature11595} {\bibfield  {journal} {\bibinfo  {journal} {Nature}\
  }\textbf {\bibinfo {volume} {491}},\ \bibinfo {pages} {570} (\bibinfo {year}
  {2012})}\BibitemShut {NoStop}%
\bibitem [{\citenamefont {Pritchard}(1983)}]{pritchard83}%
  \BibitemOpen
  \bibfield  {author} {\bibinfo {author} {\bibfnamefont {D.~E.}\ \bibnamefont
  {Pritchard}},\ }\href@noop {} {\bibfield  {journal} {\bibinfo  {journal}
  {Phys.Rev.Lett.}\ }\textbf {\bibinfo {volume} {51}},\ \bibinfo {pages} {1336}
  (\bibinfo {year} {1983})}\BibitemShut {NoStop}%
\bibitem [{\citenamefont {Newbury}\ \emph {et~al.}(1995)\citenamefont
  {Newbury}, \citenamefont {Myatt}, \citenamefont {Cornell},\ and\
  \citenamefont {Wieman}}]{Newbury95}%
  \BibitemOpen
  \bibfield  {author} {\bibinfo {author} {\bibfnamefont {N.~R.}\ \bibnamefont
  {Newbury}}, \bibinfo {author} {\bibfnamefont {C.~J.}\ \bibnamefont {Myatt}},
  \bibinfo {author} {\bibfnamefont {E.~A.}\ \bibnamefont {Cornell}}, \ and\
  \bibinfo {author} {\bibfnamefont {C.~E.}\ \bibnamefont {Wieman}},\ }\href
  {\doibase 10.1103/PhysRevLett.74.2196} {\bibfield  {journal} {\bibinfo
  {journal} {Phys. Rev. Lett.}\ }\textbf {\bibinfo {volume} {74}},\ \bibinfo
  {pages} {2196} (\bibinfo {year} {1995})}\BibitemShut {NoStop}%
\bibitem [{\citenamefont {Cirac}\ and\ \citenamefont
  {Lewenstein}(1995)}]{Cirac95}%
  \BibitemOpen
  \bibfield  {author} {\bibinfo {author} {\bibfnamefont {J.~I.}\ \bibnamefont
  {Cirac}}\ and\ \bibinfo {author} {\bibfnamefont {M.}~\bibnamefont
  {Lewenstein}},\ }\href {\doibase 10.1103/PhysRevA.52.4737} {\bibfield
  {journal} {\bibinfo  {journal} {Phys. Rev. A}\ }\textbf {\bibinfo {volume}
  {52}},\ \bibinfo {pages} {4737} (\bibinfo {year} {1995})}\BibitemShut
  {NoStop}%
\bibitem [{\citenamefont {Ovchinnikov}\ \emph {et~al.}(1997)\citenamefont
  {Ovchinnikov}, \citenamefont {Manek},\ and\ \citenamefont
  {Grimm}}]{Ovchinnikov97}%
  \BibitemOpen
  \bibfield  {author} {\bibinfo {author} {\bibfnamefont {Y.~B.}\ \bibnamefont
  {Ovchinnikov}}, \bibinfo {author} {\bibfnamefont {I.}~\bibnamefont {Manek}},
  \ and\ \bibinfo {author} {\bibfnamefont {R.}~\bibnamefont {Grimm}},\ }\href
  {\doibase 10.1103/PhysRevLett.79.2225} {\bibfield  {journal} {\bibinfo
  {journal} {Phys. Rev. Lett.}\ }\textbf {\bibinfo {volume} {79}},\ \bibinfo
  {pages} {2225} (\bibinfo {year} {1997})}\BibitemShut {NoStop}%
\bibitem [{\citenamefont {Miller}\ \emph {et~al.}(2002)\citenamefont {Miller},
  \citenamefont {D\"urr},\ and\ \citenamefont {Wieman}}]{Miller02}%
  \BibitemOpen
  \bibfield  {author} {\bibinfo {author} {\bibfnamefont {K.~W.}\ \bibnamefont
  {Miller}}, \bibinfo {author} {\bibfnamefont {S.}~\bibnamefont {D\"urr}}, \
  and\ \bibinfo {author} {\bibfnamefont {C.~E.}\ \bibnamefont {Wieman}},\
  }\href {\doibase 10.1103/PhysRevA.66.023406} {\bibfield  {journal} {\bibinfo
  {journal} {Phys. Rev. A}\ }\textbf {\bibinfo {volume} {66}},\ \bibinfo
  {pages} {023406} (\bibinfo {year} {2002})}\BibitemShut {NoStop}%
\bibitem [{\citenamefont {Janis}\ \emph {et~al.}(2005)\citenamefont {Janis},
  \citenamefont {Banks},\ and\ \citenamefont {Bigelow}}]{Janis05}%
  \BibitemOpen
  \bibfield  {author} {\bibinfo {author} {\bibfnamefont {J.}~\bibnamefont
  {Janis}}, \bibinfo {author} {\bibfnamefont {M.}~\bibnamefont {Banks}}, \ and\
  \bibinfo {author} {\bibfnamefont {N.~P.}\ \bibnamefont {Bigelow}},\ }\href
  {\doibase 10.1103/PhysRevA.71.013422} {\bibfield  {journal} {\bibinfo
  {journal} {Phys. Rev. A}\ }\textbf {\bibinfo {volume} {71}},\ \bibinfo
  {pages} {013422} (\bibinfo {year} {2005})}\BibitemShut {NoStop}%
\bibitem [{\citenamefont {Griesmaier}\ \emph {et~al.}(2009)\citenamefont
  {Griesmaier}, \citenamefont {Greiner}, \citenamefont {Sebastian},
  \citenamefont {Aghajani-Talesh}, \citenamefont {Falkenau}, \citenamefont
  {Rehme},\ and\ \citenamefont {Pfau}}]{axel09}%
  \BibitemOpen
  \bibfield  {author} {\bibinfo {author} {\bibfnamefont {A.}~\bibnamefont
  {Griesmaier}}, \bibinfo {author} {\bibfnamefont {A.}~\bibnamefont {Greiner}},
  \bibinfo {author} {\bibfnamefont {J.}~\bibnamefont {Sebastian}}, \bibinfo
  {author} {\bibfnamefont {A.}~\bibnamefont {Aghajani-Talesh}}, \bibinfo
  {author} {\bibfnamefont {M.}~\bibnamefont {Falkenau}}, \bibinfo {author}
  {\bibfnamefont {P.}~\bibnamefont {Rehme}}, \ and\ \bibinfo {author}
  {\bibfnamefont {T.}~\bibnamefont {Pfau}},\ }\href@noop {} {\bibfield
  {journal} {\bibinfo  {journal} {J. Phys B: At. Mol. Opt. Phys.}\ }\textbf
  {\bibinfo {volume} {42}},\ \bibinfo {pages} {145306} (\bibinfo {year}
  {2009})}\BibitemShut {NoStop}%
\bibitem [{\citenamefont {Aghajani-Talesh}\ \emph {et~al.}(2010)\citenamefont
  {Aghajani-Talesh}, \citenamefont {Falkenau}, \citenamefont {Volchkov},
  \citenamefont {Trafford}, \citenamefont {Pfau},\ and\ \citenamefont
  {Griesmaier}}]{anoush10}%
  \BibitemOpen
  \bibfield  {author} {\bibinfo {author} {\bibfnamefont {A.}~\bibnamefont
  {Aghajani-Talesh}}, \bibinfo {author} {\bibfnamefont {M.}~\bibnamefont
  {Falkenau}}, \bibinfo {author} {\bibfnamefont {V.}~\bibnamefont {Volchkov}},
  \bibinfo {author} {\bibfnamefont {L.}~\bibnamefont {Trafford}}, \bibinfo
  {author} {\bibfnamefont {T.}~\bibnamefont {Pfau}}, \ and\ \bibinfo {author}
  {\bibfnamefont {A.}~\bibnamefont {Griesmaier}},\ }\href@noop {} {\bibfield
  {journal} {\bibinfo  {journal} {New J. Phys}\ }\textbf {\bibinfo {volume}
  {12}},\ \bibinfo {pages} {065018} (\bibinfo {year} {2010})}\BibitemShut
  {NoStop}%
\bibitem [{\citenamefont {Greiner}\ \emph {et~al.}(2007)\citenamefont
  {Greiner}, \citenamefont {Sebastian}, \citenamefont {Rehme}, \citenamefont
  {Aghajani-Talesh}, \citenamefont {Griesmaier},\ and\ \citenamefont
  {Pfau}}]{alex07}%
  \BibitemOpen
  \bibfield  {author} {\bibinfo {author} {\bibfnamefont {A.}~\bibnamefont
  {Greiner}}, \bibinfo {author} {\bibfnamefont {J.}~\bibnamefont {Sebastian}},
  \bibinfo {author} {\bibfnamefont {P.}~\bibnamefont {Rehme}}, \bibinfo
  {author} {\bibfnamefont {A.}~\bibnamefont {Aghajani-Talesh}}, \bibinfo
  {author} {\bibfnamefont {A.}~\bibnamefont {Griesmaier}}, \ and\ \bibinfo
  {author} {\bibfnamefont {T.}~\bibnamefont {Pfau}},\ }\href {\doibase
  10.1088/0953-4075/40/5/F01} {\bibfield  {journal} {\bibinfo  {journal} {J.
  Phys B: At. Mol. Opt. Phys.}\ }\textbf {\bibinfo {volume} {40}},\ \bibinfo
  {pages} {F77} (\bibinfo {year} {2007})}\BibitemShut {NoStop}%
\bibitem [{\citenamefont {Olson}\ \emph {et~al.}(2006)\citenamefont {Olson},
  \citenamefont {Mhaskar},\ and\ \citenamefont {Raithel}}]{Olson:2006.1}%
  \BibitemOpen
  \bibfield  {author} {\bibinfo {author} {\bibfnamefont {S.~E.}\ \bibnamefont
  {Olson}}, \bibinfo {author} {\bibfnamefont {R.~R.}\ \bibnamefont {Mhaskar}},
  \ and\ \bibinfo {author} {\bibfnamefont {G.}~\bibnamefont {Raithel}},\ }\href
  {\doibase 10.1103/PhysRevA.73.033622} {\bibfield  {journal} {\bibinfo
  {journal} {Physical Review A}\ }\textbf {\bibinfo {volume} {73}},\ \bibinfo
  {pages} {033622} (\bibinfo {year} {2006})}\BibitemShut {NoStop}%
\bibitem [{\citenamefont {Cren}\ \emph {et~al.}(2002)\citenamefont {Cren},
  \citenamefont {Roos}, \citenamefont {Aclan}, \citenamefont {Dalibard},\ and\
  \citenamefont {Guery-Odelin}}]{Cren:2002.1}%
  \BibitemOpen
  \bibfield  {author} {\bibinfo {author} {\bibfnamefont {P.}~\bibnamefont
  {Cren}}, \bibinfo {author} {\bibfnamefont {C.}~\bibnamefont {Roos}}, \bibinfo
  {author} {\bibfnamefont {A.}~\bibnamefont {Aclan}}, \bibinfo {author}
  {\bibfnamefont {J.}~\bibnamefont {Dalibard}}, \ and\ \bibinfo {author}
  {\bibfnamefont {D.}~\bibnamefont {Guery-Odelin}},\ }\href {\doibase
  10.1140/epjd/e2002-00106-3} {\bibfield  {journal} {\bibinfo  {journal}
  {European Physical Journal D}\ }\textbf {\bibinfo {volume} {20}},\ \bibinfo
  {pages} {107} (\bibinfo {year} {2002})}\BibitemShut {NoStop}%
\bibitem [{\citenamefont {Falkenau}\ \emph {et~al.}(2012)\citenamefont
  {Falkenau}, \citenamefont {Volchkov}, \citenamefont {R\"uhrig}, \citenamefont
  {Gorniaczyk},\ and\ \citenamefont {Griesmaier}}]{falkenau12}%
  \BibitemOpen
  \bibfield  {author} {\bibinfo {author} {\bibfnamefont {M.}~\bibnamefont
  {Falkenau}}, \bibinfo {author} {\bibfnamefont {V.~V.}\ \bibnamefont
  {Volchkov}}, \bibinfo {author} {\bibfnamefont {J.}~\bibnamefont {R\"uhrig}},
  \bibinfo {author} {\bibfnamefont {H.}~\bibnamefont {Gorniaczyk}}, \ and\
  \bibinfo {author} {\bibfnamefont {A.}~\bibnamefont {Griesmaier}},\ }\href
  {\doibase 10.1103/PhysRevA.85.023412} {\bibfield  {journal} {\bibinfo
  {journal} {Phys. Rev. A}\ }\textbf {\bibinfo {volume} {85}},\ \bibinfo
  {pages} {023412} (\bibinfo {year} {2012})}\BibitemShut {NoStop}%
\bibitem [{\citenamefont {Castin}\ \emph {et~al.}(1998)\citenamefont {Castin},
  \citenamefont {Cirac},\ and\ \citenamefont {Lewenstein}}]{Lewenstein98}%
  \BibitemOpen
  \bibfield  {author} {\bibinfo {author} {\bibfnamefont {Y.}~\bibnamefont
  {Castin}}, \bibinfo {author} {\bibfnamefont {J.~I.}\ \bibnamefont {Cirac}}, \
  and\ \bibinfo {author} {\bibfnamefont {M.}~\bibnamefont {Lewenstein}},\
  }\href {\doibase 10.1103/PhysRevLett.80.5305} {\bibfield  {journal} {\bibinfo
   {journal} {Phys. Rev. Lett.}\ }\textbf {\bibinfo {volume} {80}},\ \bibinfo
  {pages} {5305} (\bibinfo {year} {1998})}\BibitemShut {NoStop}%
\bibitem [{Note1()}]{Note1}%
  \BibitemOpen
  \bibinfo {note} {One can estimate from Fig. \ref {fig:RFamplitude} that the
  amount of scattered light from loading and Sisyphus cooling are approximately
  the same.}\BibitemShut {Stop}%
\bibitem [{Note2()}]{Note2}%
  \BibitemOpen
  \bibinfo {note} {The only constraint for this scheme is that the longitudinal
  velocity of the beam has to match the height of the barrier. In Sec. \ref
  {sec:noevap} this constraint is further relaxed.}\BibitemShut {Stop}%
\bibitem [{\citenamefont {Roos}\ \emph {et~al.}(2003)\citenamefont {Roos},
  \citenamefont {Cren}, \citenamefont {Guéry-Odelin},\ and\ \citenamefont
  {Dalibard}}]{Roos:2003.2}%
  \BibitemOpen
  \bibfield  {author} {\bibinfo {author} {\bibfnamefont {C.~F.}\ \bibnamefont
  {Roos}}, \bibinfo {author} {\bibfnamefont {P.}~\bibnamefont {Cren}}, \bibinfo
  {author} {\bibfnamefont {D.}~\bibnamefont {Guéry-Odelin}}, \ and\ \bibinfo
  {author} {\bibfnamefont {J.}~\bibnamefont {Dalibard}},\ }\href@noop {}
  {\bibfield  {journal} {\bibinfo  {journal} {Europhysics Letters}\ }\textbf
  {\bibinfo {volume} {61}},\ \bibinfo {pages} {187} (\bibinfo {year}
  {2003})}\BibitemShut {NoStop}%
\end{thebibliography}

%

\end{document}